\documentclass[natbib]{svjour3}                     
\smartqed  
\usepackage{graphicx}
 \usepackage{aps-bibstyle}  
\begin{document}

\title{Some Uncomfortable Thoughts on the Nature of Gravity, Cosmology, and 
the Early Universe
\thanks{Contribution to the `Nature of Gravity' conference at the 
International Space Science Institute, October 2008, Bern, Switzerland.}
}

\author{L P Grishchuk}


\institute{L P Grishchuk \at School of Physics and Astronomy, 
Cardiff University \\ Cardiff CF234AA, 
United Kingdom\\ and \\ Sternberg Astronomical Institute, Moscow State
University\\ Moscow 119899, Russia\\
              Tel.: +44 29 20874665\\
              Fax: +44 20 20874056\\
              \email{grishchuk@astro.cf.ac.uk}           
}

\date{Received: date / Accepted: date}

\maketitle

\begin{abstract}

A specific theoretical framework is important for designing and conducting an 
experiment, and for interpretation of its results. The field of gravitational 
physics is expanding, and more clarity is needed. It appears that some
popular notions, such as `inflation' and `gravity is geometry', have become
more like liabilities than assets. A critical analysis is presented and the 
ways out of the difficulties are proposed. 


\end{abstract}

\section{Introduction}
\label{intro}

The proximity of the site of this gravitational conference to CERN and its 
recently completed Large Hadron Collider (LHC) reminds us of the affinity of 
our ultimate goals in the study of micro- and macro-worlds. Although the LHC 
will be investigating the minute particles - hadrons, the outreach page of the 
LHC web-site explains to the wide public that the ``aim of the exercise is to 
smash protons...into each other and so recreate conditions a fraction of a 
second after the big bang" [\cite{lhc}]. One can also see clarifications in 
other 
sources of information, according to which the ``LHC experiments...will probe 
matter as it existed at the very beginning of time", and that this is 
a ``new era of understanding about the origins and evolution of the universe". 

These cosmological, rather than particle-physical, explanations can perhaps 
be justified, at least in part, by the alleged difficulties that our 
colleagues encountered in communication with general public. According to a
circulating rumor, at the time when the opening of the LHC was widely covered 
by TV media, the BBC received numerous messages from angry parents 
who complained that ``you keep talking about all these hardons while children 
may be watching". Surely, the descriptive cosmology is safer and easier to 
convey to the public than the notions of high-energy physics. But in the long 
run, it is indeed true that the fundamental question of the birth of the 
Universe is always in the background of our research intentions. This problem 
fascinates both scientific communities, as well as some part of the rest of 
population.

Undoubtedly, new important discoveries will be made at LHC and they will 
bring us closer to answers to very deep issues in physics. However, 
the record-breaking energies of $10^{4}{\rm GeV}$ at our accelerators are still 
very far away from energies we need to understand in 
order to tackle the problem of the origins of the Universe. The ability of LHC 
to answer the questions on the physics of the very early Universe at 
$(10^{15} -10^{19}){\rm GeV}$ can be compared with the ability of a telescope 
hardly resolving a planetary system to answer the questions on the structure 
of a human hair. In these matters, our best hopes are associated with
cosmic rather than laboratory studies. It is now recognized that 
there is no better way of trying to unveil the origins of the Universe than by 
measuring the relic gravitational waves that were generated by a strong 
variable gravitational field of the emerging Universe. We may be close to
discovering relic gravitational waves, and thus answering some of the most
fundamental questions. These efforts, as well as some principal issues on the
nature of gravity, will be discussed below.

\section{Spontaneous Birth of the Universe}
\label{sec:1}

The question of origins arises inevitably given the nonstationary character of
the world that we observe at large scales. The broad-brush picture of the 
expanding homogeneous and isotropic distribution of matter and fields is pretty 
accurate as a zero-order approximation. We believe that this approximation was 
even better in the past, because the existing deviations from homogeneity and
isotropy seem to have been growing in the course of time. It is known from 
observations that the present size $l_p$ of our patch of approximate 
large-scale homogeneity and isotropy is at least as big as the present-day 
Hubble radius $l_H = c/H_0 \approx 10^{28}{\rm cm}$. Combining available 
observations with plausible statistical assumptions one can conclude that 
the size of this patch is significantly bigger than $l_H$. It was evaluated to 
be about $500 l_H$ [\cite{gr92}] and maybe larger. In the present 
discussion, we will limit ourselves with $l_p = 10^{3} l_H$. It is also 
known from observations that the present averaged energy density 
$\rho_{p} c^2$ of all sorts of matter in our patch is close to 
$\rho_p =3 H_0^2/8 \pi G \approx 10^{-29} {\rm g}/{\rm cm}^3$.

Given the observed expansion, one can extrapolate $\rho(t)$ and $l(t)$ back in 
time by the known laws of physics. Temporarily leaving aside the possible 
recent interval of accelerated evolution governed by ``dark energy" 
(see Sec.\ref{sec:6}), we will first have $\rho(t) \propto 1/l^{3} (t)$ 
at the matter-dominated stage, up to $\rho_{eq} 
\approx 10^{-20}{\rm g}/{\rm cm}^3$, and then  $\rho(t) \propto 1/l^{4}(t)$ 
at the radiation-dominated stage. Eventually one encounters a cosmological 
singularity characterized by the infinitely large energy density. It is 
reasonable to think that there must exist a smarter answer to the question 
of initial state of the observed Universe than singularity. Singularity is 
likely to be only a sign of breakdown of the currently available theories. 

\begin{figure}
\includegraphics[width=12cm,height=10cm]{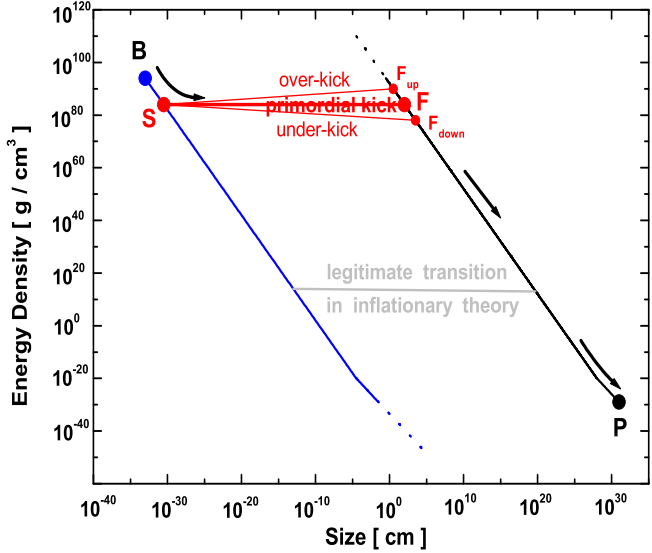}
\caption{A primordial kick (red lines) is required in order to reach the 
present state of the Universe P from the birth event B.}
\label{birth}  
\end{figure}

It is tempting to begin from the limits of applicability of current 
theories, that is, from the Planck density 
$\rho_{Pl} = c^5/G^2 \hbar \approx 10^{94}{\rm g}/{\rm cm}^{3}$ 
and the Planck size $l_{Pl} = (G \hbar/c^3)^{1/2} \approx 10^{-33}{\rm cm}$, 
and 
imagine that the embryo Universe was somehow created by a quantum-gravity (or
by a `theory-of-everything') process in this state and then started to expand 
(see [\cite{zeld86}] and references there). Conceptually, it is easier to 
imagine that the created Universe was spatially closed, which means that
its total energy, including gravity, was zero and remains zero. Also, the 
concept of a closed universe helps avoid the question of environment. But a 
closed universe is not strictly necessary for our line of argument, so one 
can think of a small configuration in a possibly larger system. It is also 
plausible to think that the classical evolution of our patch of the Universe 
could have started with a natal size somewhat larger than $l_{Pl}$ and a natal 
energy density, of whatever matter that was there, somewhat lower than 
$\rho_{Pl}$. 

The trouble is, however, that the very natural hypothesis of spontaneous 
birth of the observed Universe will not bring us anywhere near our present 
state characterized by $\rho_p$ and $l_p$, unless we make additional 
assumptions [\cite{zeld86}]. 

In Fig.\ref{birth}, the present state of the accessible Universe is marked by 
the point P. The hypothesized birth of the Universe is marked by the point B.
If we go in the past from the point P according to the laws of matter-dominated
and radiation-dominated evolutions (black curve), we totally miss the desired 
point B. Indeed, the energy density reaches the Planck value $\rho_{Pl}$ when 
the size of the model universe is $0.3{\rm cm}$ instead of the required 
$10^{-33}{\rm cm}$. On the other hand, if we descend from the point B according 
to the laws of radiation-dominated and then matter-dominated evolutions 
(blue curve), we totally miss the point P. Indeed, the present energy density 
$\rho_p$ is reached while the size of the model universe is $0.03{\rm cm}$ 
instead of the required $10^{31}{\rm cm}$. And the present size 
$10^{31}{\rm cm}$ is reached when the density drops to 
$10^{-126}{\rm g}/{\rm cm}^3$ instead of the required $\rho_p$.
If the size $l_p$ of the homogeneous isotropic patch is larger than the 
assumed $l_p = 10^{3} l_H$, the point P and the black curve in Fig.\ref{birth} 
move to the right, and the mis-match of the curves is only exacerbated.  

The only way to reach P from B is to assume that the newly born Universe
has experienced a `primordial kick' allowing the point of evolution to 
jump over from the blue curve to the black curve. During the kick, the size of 
the Universe (or, better to say, the size of the patch of homogeneity and 
isotropy) should increase by about 33 orders of magnitude, but the energy 
density may not change too much or may stay constant. 
In the zero-order approximation, when the homogeneity and isotropy of the 
patch are maintained, the level of the energy density and the specific points 
of the start S and finish F of the kick transition are not very important,
as we can reach P from B by many ways. However, the actual route becomes 
extremely important in the first-order approximation, when we have to take 
into account the quantum-mechanical generation of primordial cosmological 
perturbations. 

If we are on the right track with this whole picture, the observations 
indicate (see below) that the actual kick trajectory took place at the energy 
densities around $10^{-10} \rho_{Pl}$ when the Hubble parameter $H$ was
around $H_i =10^{-5} H_{Pl} = 10^{-5}/ t_{Pl}$ (horizontal red line in 
Fig.\ref{birth}), with some possibility of a slight `over-kick' to point 
F$_{\rm up}$ or `under-kick' to point F$_{\rm down}$, as shown by tilted 
red lines in Fig.\ref{birth}.

The general relativity allows the required kick trajectories, but only if the 
properties of the primeval matter were not like those that we deal with in 
laboratories (even at LHC). The energy density $\rho c^2$ and the effective 
pressure $p$ should satisfy the condition $\rho c^2 + p =0$ for the 
kick to proceed with constant energy 
density, and $\rho c^2 + p <0$ for over-kick with increasing energy density 
or $\rho c^2 + p >0$ for under-kick with decreasing energy density. The 
kick Hubble parameter $H_{i}(t)$ remains constant ($\dot{H} = 0$) for the 
$\rho c^2 + p =0$ case, and it is slightly increasing ($\dot{H} >0$) or 
decreasing ($\dot{H} <0$) for over-kick and under-kick trajectories, 
respectively. If the existing indications (see below) that the initial kick 
did indeed happen are further supported by observations, the detailed 
inquiry in the properties of the substance that might have driven the 
kick will be paramount.

Solutions with $\rho c^2 + p \geq 0$ and $p \approx - \rho c^2$ are allowed by
some versions of the scalar field within general relativity. They are usually
associated with the notion of inflation (see, for example, [\cite{wein}]). The 
case with $p= - \rho c^2$ is known as the standard (de Sitter) inflation. 
The scalar field plus gravity model may not be what we are searching  
for, but at least it was shown [\cite{bgkz}] that the required solutions 
with $p \approx -\rho c^2$ are attractors in the space of all solutions of 
this dynamical system, so these are typical solutions which can bring us 
from the quantum boundary S [\cite{bgkz}] to the end point 
F$_{\rm down}$. (Scalar field models do not admit the over-kick trajectories.)

It is often stated that inflation was invented by particle physicists 
in order to solve outstanding cosmological problems. The length of the list 
of solved problems depends on the enthusiasm of the writing 
inflationary author. To me personally, most of these problems and solutions 
smack of the Soviet-style management, when the goverment creates a problem 
and then demands the credit for heroically solving it. In any case, 
inflation was always assumed to have lasted sufficiently long; otherwise the 
point of evolution would have fell short of reaching the point 
F$_{\rm down}$ on the black curve in Fig.\ref{birth}. Whatever the problems 
the inflationary hypothesis solves, they are automatically solved by the 
hypothesis of primordial kick, whose only rationale is to serve as
an `umbilical cord' facilitating the joining of the birth event B with the 
present state of the Universe P.

While in the approximation of homogeneity and isotropy the hypothesis of
inflation and the hypothesis of initial kick are equal, in the sense that
they both solve problems by making plausible assumptions, at the level of 
cosmological perturbations the inflationary theory got everything wrong, 
as we shall see below, in Sec.\ref{sec:3}.

\section{Primordial Cosmological Perturbations}
\label{sec:2}

The essentially classical and highly symmetric, i.e. homogeneous and 
isotropic, solution describing the initial kick should be augmented by quantum 
fluctuations of the fields that were present there. It is natural to assume 
that these quantized fields were initially in their ground (vacuum) states. 
If the field is properly, superadiabatically, coupled to the strong 
gravitational field of the kick solution, the quantum-mechanical Schr\"odinger 
evolution transforms the initial vacuum state of the field into a multiparticle 
(strongly squeezed vacuum) state. This is called the superadiabatic, 
or parametric, amplification; for a recent review of the subject, see 
[\cite{gr07}]. Not all fields couple properly to the highly symmetric 
solutions under discussion. For example, electromagnetic 
fields do not. But the gravitational field perturbations do 
couple superadiabatically to the gravitational field of the kick, and 
therefore they can be amplified. Apparently, this is how the present-time 
complexity arises from the past-time simplicity.     

The chief gravitational field perturbations are two transverse-traceless
degrees of freedom describing gravitational waves and two degrees of 
freedom, scalar and \\
longitudinal-longitudinal one, which in general relativity 
exist only if they are accompanied by perturbations of matter. These latter 
two degrees of freedom describe density perturbations, and usually only one 
of them is independent. For models of matter such as scalar fields, the 
equations for density perturbations are almost identical to the equations for 
gravitational waves. The arising multiparticle states at each frequency  
are conveniently combined in the power spectra of these quantum-mechanically 
generated perturbations. The crucial quantity, both for gravitational
waves and density perturbations, is the primordial power spectrum of the 
gravitational field (metric) perturbations. This is because the gravitational 
field perturbations survive numerous transformations of the matter content of
the Universe at the end of the kick and later, and therefore they 
provide the unambiguous input values for physics and calculations at the 
radiation-dominated and matter-dominated stages. 

The values of the Hubble parameter along the kick trajectory and the 
shape of the kick trajectory itself are very important, because they define 
the numerical levels and spectral slopes of the primordial metric spectra. The 
Hubble parameter $H_i$ determines the amplitude of metric perturbations, 
whereas the tilt of the red lines in Fig.\ref{birth} determines the tilt 
of the spectrum. The primordial metric power spectrum $P(n)$ is a 
function of the time-independent wavenumber $n$, where the wavenumber 
$n_H=4\pi$ corresponds to the wavelength which will be equal 
to $l_H$ today. 

\begin{figure}
\begin{center}
\includegraphics[width=10cm,height=8cm]{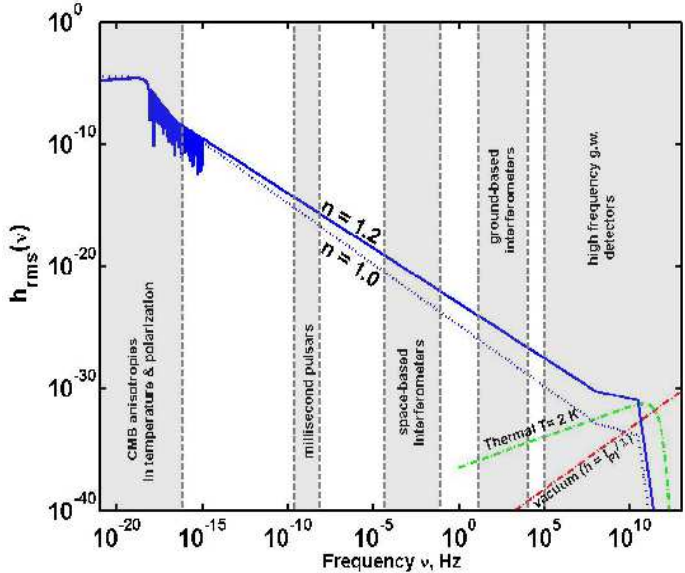}
\end{center}
\caption{The present-day spectrum of the root-mean-square amplitude 
$h_{rms}(\nu)$ of relic gravitational waves. The solid line corresponds 
to the primordial spectral index ${\rm n}_t =0.2$ (${\rm n} =1.2$), while 
the dashed line is for ${\rm n}_t =0$ (${\rm n} =1.0$).}
\label{gwSpectrum}
\end{figure}

Very similar forms of equations for gravitational waves (gw) and 
density perturbations (dp), and identically the same physics of 
their superadiabatic amplification, translate into very similar power spectra 
$P(n)$. In a good approximation, sufficient for our purposes, the generated 
gw and dp power spectra can be written in the power-law forms
\begin{equation}
\label{psa}
P(n)~(gw) = \left(\frac{H_i}{H_{Pl}}\right)^2 
\left(\frac{n}{n_H}\right)^{{\rm n}_t}, ~~~~~~~~~~~
P(n)~(dp) = \left(\frac{H_i}{H_{Pl}}\right)^2 
\left(\frac{n}{n_H}\right)^{{\rm n}_s -1},
\end{equation}
where the spectral indices ${\rm n}_t,~{\rm n}_s$ are constants, and 
${\rm n}_s -1 \approx {\rm n}_t $. For flat kick evolutions, 
i.e. for horizontal lines like the red line in Fig.\ref{birth}, 
the generated spectra have flat (Harrison-Zeldovich-Peebles) 
shape, i.e. ${\rm n}_t = {\rm n}_s -1 = 0$. The over-kick or under-kick 
evolutions tilt the spectrum toward the `blue' or `red' shapes, 
respectively. Of course, it is simplistic to expect that the red 
transitions in Fig.\ref{birth} should be strictly straight lines and the 
spectral indices ${\rm n}_t,~{\rm n}_s$ strict numbers independent of the 
wavenumber $n$. In simple models driven by scalar fields, the spectral 
indices are actually slowly decreasing functions of $n$. 

The derivation of primordial spectra is based only on general relativity and
quantum mechanics. The comparison of spectra and their consequences with 
observations is our best chance to learn whether the kick did indeed take 
place and how it looked like. In particular, assuming that the observed 
anisotropies of the cosmic microwave background radiation (CMB) are 
indeed caused by cosmological perturbations of quantum-mechanical origin, 
we conclude that the kick Hubble parameter $H_i$ was close 
to $10^{-5} H_{Pl}$, because this kick has generated the amplitudes 
of metric perturbations at the level of $10^{-5}$, which in their turn 
produced the observed large-scale CMB anisotropy at the level of $10^{-5}$. 

The perturbations with wavelengths much shorter than the radius $l_p$ of 
our patch were processed in the course of evolution, and therefore the form 
of the spectrum in the recombination era (and today) differs from primordial. 
The spectrum is no longer a smooth function of the wavenumber (or frequency) 
but contains maxima and minima. This is a consequence of the standing-wave 
pattern of the primordial metric perturbations - the inevitable feature of 
the underlying quantum-mechanical squeezing. The power-law slope of the 
envelope of the oscillations also changes. 

As an example, in Fig.\ref{gwSpectrum}, taken from [\cite{gr07}], we show 
today's power spectra as functions of frequency $\nu$ for relic gravitational 
waves normalized to the observed CMB anisotropies. Only a few first cycles 
of oscillations at lowest frequencies are shown. Two possibilities are 
depicted in this figure: the dashed line is the resulting spectrum 
which originated in the superadiabatic amplification during the 
horizontal red line transition of Fig.\ref{birth}, whereas the solid line 
is for the resulting spectrum originated in a slightly over-kick transition. 
At the highest relevant frequencies, before the spectrum sharply goes
down, one can see a noticable increase of power in $h_{rms}$. This is the 
result of the after-kick piece of evolution governed by matter with a 
stiff (Zeldovich) equation of state $p=  \rho c^2$. This piece of evolution 
was assumed in this calculation, but it is not guaranteed. We may be not so 
lucky to have to deal in the future high-frequency experiments with this 
increased gravitational-wave power.

\section{Cosmological Perturbations in Inflationary Theory}
\label{sec:3}

Now we turn to what is broadly addressed as inflation, and specifically 
to predictions of inflationary theory on cosmological perturbations. 
Predictions of inflationary theory are dramatically different from what 
was described above, in Sec.\ref{sec:2}. In attempt of deriving density 
perturbations, inflationary authors invariably begin with vacuum 
fluctuations of the scalar field (inflaton) in de Sitter space-time. Then, 
after some jumping between variables and gauges, they conclude that 
the amplitudes of scalar metric perturbations (often denoted 
by letters $\zeta$ or $\cal R$) should be infinitely large, in the same 
physical system and right from the very beginning. Instead of Eq.(\ref{psa}), 
the inflation-predicted primordial metric power spectrum for density 
perturbations reads:
\begin{equation}
\label{psain}
P(n)~(dp)|_{\rm inflation} =  \frac{1}{\epsilon} \left(\frac{H_i}{H_{Pl}}
\right)^2 \left(\frac{n}{n_H}\right)^{{\rm n}_s -1},
\end{equation} 
where the parameter $\epsilon$ is $\epsilon \equiv - \frac{{\dot H}}{H^2}$. 
This parameter is almost a constant in cases under discussion. The standard 
(de Sitter) inflation is characterized by $\epsilon =0$ ($p= - \rho c^2$), and 
therefore the inflation-predicted standard spectrum blows up to infinity at 
every wavelength and for any non-zero value of $H_i$. 

Formula (\ref{psain}) is the main contribution of inflationary theory
to the subject of cosmological perturbations. Having arrived at the incorrect 
formula with the arbitrarily small factor $\epsilon$ in the denominator of 
this formula, inflationary theory elevates the question of the ``energy scale 
of inflation" to the status of a major scientific problem, which 
inflationists will be happily solving for decades to come. Indeed, one is now 
free to start with energy densities $\rho_i$, say, 80 orders of magnitude 
smaller than $\rho_{Pl}$ and the Hubble parameter $H_i$ 40 orders of 
magnitude smaller than $H_{Pl}$ (grey line in Fig.\ref{birth}). One can still 
claim that (s)he has built a successful inflationary model, because the 
required level $10^{-5}$ of primordial metric amplitudes for density 
perturbations can now be achieved by simply making the grey line sufficiently 
horizontal (but not exactly horizontal) and thus making the denominator 
$\epsilon$ in Eq.(\ref{psain}) sufficiently small (but not exactly zero). 
Astrophysics does not permit one to take $H_i$ smaller than 
the value of $H$ in the era of nucleosynthesis, i.e. below the grey line 
in Fig.\ref{birth}. Otherwise, one could have started with the ``energy 
scale of inflation" equal to the energy density of water and still build a 
successful inflationary cosmology.

Inflationary theory substitutes the predicted divergency of density 
perturbations at $\epsilon =0$, Eq.(\ref{psain}), by the claim that it is 
the amount of relic gravitational waves that should be small. This is 
meant to be represented by the ``tensor-to-scalar" ratio $r$, which is the 
ratio of the gw power spectrum from Eq.(\ref{psa}) to the dp power spectrum 
from Eq.(\ref{psain}). In inflationary theory, it is written as
\begin{equation}
\label{tsr}
r= 16 \epsilon = - 8{\rm n}_t.
\end{equation}

The WMAP (Wilkinson Microwave Anisotropy Probe) team reports the limits on 
the parameter $r$ that were found from the observations [\cite{koma}]. One can 
see these limits in the form of the likelihood function for $r$ in the left 
panel of Fig. 3 in Ref.\cite{koma}. The maximum of the likelihood function
is at $r=0$, which means that the most likely value of the parameter $r$ 
is $r=0$. Then, according to Eq.(\ref{tsr}), the most likely value of 
$\epsilon$ is $\epsilon =0$. Since $\epsilon$ is in the denominator of the
inflation-predicted power spectrum (\ref{psain}), one concludes that if 
the inflationary predictions are correct, then the most likely values of 
density perturbations responsible for the data observed and analyzed by the 
WMAP team are infinitely large. Nevertheless, this situation is often 
qualified as the ``striking success of inflation" [\cite{baum}] demonstrated 
by the WMAP findings.

For the parameter $r$, the inflationary theory predicts practically 
everything what one can possibly imagine, including something like 
$r \leq 10^{-24}$ as the inflationary outcome, based on Eq.(\ref{psain}), 
of the most sophisticated string-inspired models (for a review, see 
[\cite{baum}] and references there). The big problem is not that the seemingly 
most advanced theories predict $r \leq 10^{-24}$, thus making the
search for relic gravitational waves look ridiculous. The big problem is that 
when the relic gravitational waves are discovered, the proposers, being 
misguided by their own derivations, will reject microphysical theories 
which in fact may be perfectly viable, and will accept theories which in 
fact may be totally wrong.

\section{Discovering Relic Gravitational Waves in the Cosmic Microwave 
Background Radiation}
\label{sec:4}

There are several reasons to believe that the observed CMB anisotropies are 
caused by cosmological perturbations of quantum-mechanical origin. Probably 
the major one is the observed oscillations in the CMB power spectra, which 
are likely to be a reflection of the quantum-mechanical squeezing and the 
associated standing-wave pattern of the primordial metric perturbations. 
If cosmological perturbations do indeed have quantum-mechanical origin, 
Eq.(\ref{psa}) implies that the gw and dp contributions to the 
lower-order CMB multipoles $\ell$ should be of the same order of magnitude.
Relic gravitational waves have not been scattered or absorbed in any 
significant amount since the time of their quantum-mechanical generation. 
The explicit identification of relic gravitational waves in the data would 
be a monumental step in the study of the primordial kick and the origins of 
the Universe. 

Density perturbations and gravitational waves affect temperature and 
polarization of the CMB. The polarization is usually characterized by the 
two components - E and B. The B component is not generated by density 
perturbations, and therefore it is often stated that the aim of the 
exercise is to detect B-modes. This is not so. The aim of the exercise is 
to detect relic gravitational waves, not to detect B-modes. Gravitational 
waves are present in all correlation functions of the CMB, and one should be 
smart enough to distinguish their contribution from competing contributions. 
The B-mode channel of information has some advantages, but many disadvantages 
too. For example, the 5-year WMAP data on the BB correlation function are not 
informative because they are mostly noise. At the same time, the 
much stronger TE signal is recorded at a large number of data points, which 
allows one to derive meaningful conclusions about the gw contribution.  
\begin{figure}
\begin{center}
\includegraphics[width=13cm,height=11cm]{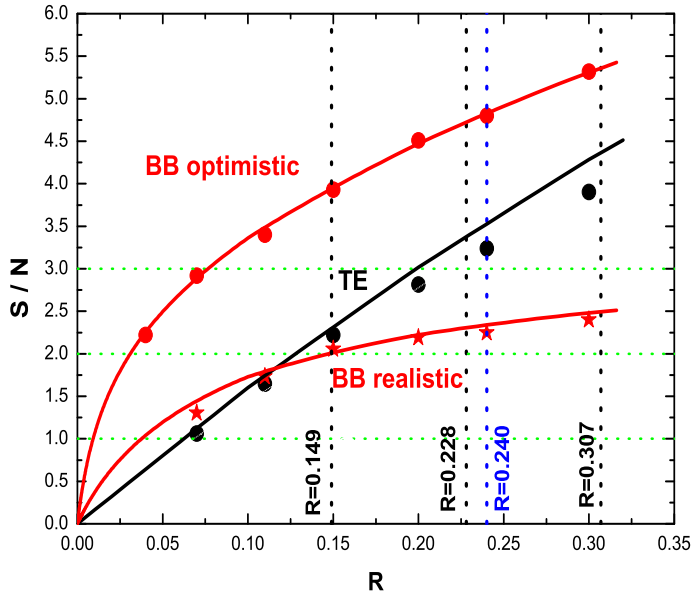}
\end{center}
\caption{The signal to noise ratio $S/N$ as a function of $R$
for the $TE$ (black) and $BB$ (red) observational channels.
The points show numerical results, whereas the curves are analytical
approximations.}\label{section6_fig3}
\end{figure}

The 5-year WMAP TE data were thorougly analyzed in Ref.\cite{zbg}. The
conclusion of this investigation is such that the lower-$\ell$ TE and TT data 
do contain a hint of presence of the gravitational wave contribution. In terms
of the parameter $R$, which gives the ratio of contributions of gw and dp to 
the temperature quadrupole, the best-fit model produced $R= 0.24$.
This means that 20\% of the temperature quadrupole is accounted for by 
gravitational waves and 80\% by density perturbations. The residual WMAP 
noise is high, so the uncertainty of this determination is large, and it easily 
includes the hypothesis that there is no gw contribution at all. However, 
the uncertainty will be much smaller in the forthcoming more sensitive 
observations, most notably with the $Planck$ satellite. It is likely that 
the $Planck$ mission will be capable of strengthening our belief that the 
primordial kick did take place, and at the near-Planckian values of the 
Hubble parameter, $H_i \approx 10^{-5} H_{Pl}$.

The future TE and BB data from $Planck$ satellite were simulated and analyzed 
in [\cite{zbg}]. The quality of the future performance of the BB channel is not 
clear, so the authors discuss the `optimistic' and `realistic' options. The 
result of the analysis is shown in Fig.\ref{section6_fig3}, where the signal 
to noise parameter $S/N$ is plotted as a function of $R$. It is seen from 
the graph that if the maximum likelihood value $R= 0.24$ 
derived from the WMAP5 TE and TT data is taken as 
a real signal, it will be present at a better than 3$\sigma$ level in the 
$Planck's$ TE observational channel, and at a better than 2$\sigma$ level in 
the `realistic' BB channel. The time of discovery of relic gravitational 
waves may be nearer than is usually believed.

\section{Field-Theoretical Formulation of General Relativity}
\label{sec:5}

One of important premises in the above conclusions is the assumption that the
general relativity remains valid up to energy densities approaching
Planckian limit. Although there is no obvious reason to doubt this, any piece
of new information on the domain of applicability of a fundamental theory is
useful. The tests of gravitational theories in various circumstances and
conditions is an important part of current research, including the results 
and plans discussed at this conference. As usual, the set-up of an experiment 
and interpretaion of its results partially depend on the accepted theoretical 
framework, which in the case of general relativity seems to be obvious and well
established: ``gravity is geometry". I personally feel that the emphasis on 
the geometrical aspect of gravity has exhausted its usefulness and has 
become an impeding rather than a driving factor in gravitational physics.  

There seems to be something odd in the conviction of a large part of 
our community that the gravitational waves are ``oscillations 
in the fabric of space-time", and that the ongoing and planned 
gravitational-wave experiments are attempting to measure the ``strain in the 
fabric of space-time". This understanding is usually presented as a  
consequence of the equivalence principle which forbids, so is believed, such 
things as rigorously defined local gravitational energy density and flux of
energy. At best, this understanding allows surrogates, such as the averaged 
pseudotensors. The argument seems to be especially appropriate for the 
current research, wherein the gravitational-wave detectors are usually 
small and, in a sense, local. The size of the detector is much smaller 
than the wavelength which the detector is most sensitive to, so the 
gravitational-wave flux through the detector is supposed to be completely 
or partially removable by a coordinate transformation. (In fact, the four
pseudotensor components describing energy density and flux are removable 
by four coordinate transformation functions everywhere, not only locally.) 
Therefore, the argument goes, it is the ``squeezing and stretching of 
space-time itself" that is a proper description of the phenomenon, and not 
the absorption of the part of the gravitational wave flux by the detector, 
which makes the detector `klick' and register the event.

The problem appears to be more than simply linguistic. The concept ``gravity 
is geometry" can be misleading in the analysis of the detector's response 
and in the interpretation of results. More importantly,
it obscures the deeper understanding of gravity. Namely, the fact that the 
Einstein's general relativity is a perfectly consistent non-linear 
gravitational field theory in flat (Minkowski) space-time, with rigorously 
defined gravitational energy-momentum tensor, and with no need for geometrical 
notions of curved space-time. 

The field-theoretical approach to general relativity is based on the concept 
of gravitational field $h^{\mu \nu}(x^{\alpha})$ defined in flat 
space-time with the metric tensor $\gamma_{\mu \nu}(x^{\alpha})$. The 
curvature tensor constructed from $\gamma_{\mu \nu}(x^{\alpha})$ is 
identically zero: 
\begin{equation}
\label{R}
\breve R_{\alpha\beta\mu\nu}(\gamma_{\rho\sigma}) = 0. 
\end{equation}
The choice of coordinates $x^{\alpha}$ is in one's hands, so
the $\gamma_{\mu \nu}(x^{\alpha})$ can always be transformed, if necessary, 
to the Minkowski matrix $\eta_{\mu \nu}$ (class of Lorentzian coordinates). 
In arbitrary curvilinear coordinates, the covariant derivatives are denoted by 
a semicolon~$``;"$. 

The gravitational Lagrangian $L^{g}$ is a quadratic function of first 
derivatives of the field $h^{\mu \nu}$ (see [\cite{bg}] and references there). 
The variation of $L^{g}$ with respect to the field variables $h^{\mu \nu}$ 
leads to the gravitational equations of motion:
\begin{equation} 
\label{evac}
\frac{1}{2} \left[(\gamma^{\mu \nu} +h^{\mu\nu})(\gamma^{\alpha\beta}+
h^{\alpha\beta}) - (\gamma^{\mu \alpha} +h^{\mu\alpha})(\gamma^{\nu\beta}+
h^{\nu\beta})\right]_{; \alpha ; \beta} = \kappa t^{\mu\nu}.
\end{equation}  
In the right hand side (r.h.s) of these equations stands the gravitational 
energy-momentum tensor $t^{\mu\nu}$ (and $\kappa \equiv 8 \pi G/c^4$ ). 
The $t^{\mu\nu}$ is rigorously defined as variational derivative of $L^{g}$ 
with respect to variations of the metric tensor $\gamma_{\mu \nu}$, with the 
constraint (\ref{R}) properly taken into account. The $t^{\mu\nu}$ contains 
squares of first derivatives of the field $h^{\mu \nu}$, but not higher 
derivatives. The gravitational field equations were deliberately rearranged to
the form of Eq.(\ref{evac}), where $t^{\mu\nu}$ is the manifest source for 
the generalised wave (d'Alembert) operator standing in the left hand side 
(l.h.s.) of the equations.  

In the presence of matter Lagrangian $L^{m}$, the right hand side of
Eq.(\ref{evac}) changes:
\begin{equation} 
\label{emat}
\frac{1}{2} \left[(\gamma^{\mu \nu} +h^{\mu\nu})(\gamma^{\alpha\beta}+
h^{\alpha\beta}) - (\gamma^{\mu \alpha} +h^{\mu\alpha})(\gamma^{\nu\beta}+
h^{\nu\beta})\right]_{; \alpha ; \beta}=\kappa(t^{\mu\nu}|_{m}+\tau^{\mu\nu}).
\end{equation}  
The $\tau^{\mu\nu}$ is the matter energy-momentum tensor derived as  
variational derivative of $L^{m}$ with respect to the metric tensor 
$\gamma_{\mu \nu}$, whereas $t^{\mu\nu}|_{m}$ is the modified version of the 
gravitational energy momentum-tensor $t^{\mu\nu}$. The $t^{\mu\nu}|_{m}$ 
includes the term arising from $L^{m}$ and describing the interaction of 
gravity with matter.

The divergence (contracted covariant derivative) of the left hand side 
of equations (\ref{evac}), (\ref{emat}) vanishes identically. This means that 
the equations of motion contain the differential conservation laws 
\begin{equation}
\label{cons}
t^{\mu\nu}_{~~;\nu} =0, ~~~~~~~~   (t^{\mu\nu}|_{m}+\tau^{\mu\nu})_{;\nu} =0.
\end{equation}
The differential conservation laws can be converted into conserved integrals 
for isolated non-radiating systems. 

The geometrical description of gravity can be introduced as follows. The 
special form of $L^{g}$, which is the field-theoretical analog of the 
Hilbert-Einstein Lagrangian, and the universal coupling of gravity to matter 
assumed in $L^{m}$, which is a realization of the equivalence principle, allow 
one to `glue together' the metric tensor $\gamma^{\mu \nu}$ and the field 
variables $h^{\mu \nu}$ into one tensorial object $g^{\mu \nu}$. The 
Lagrangians $L^{g}$ and $L^{m}$, as well as the gravitational and matter 
field equations, can now be rewritten in terms of this object and its 
derivatives alone, plus matter variables. 

Specifically, one introduces $g^{\mu \nu}$ according to the rule   
\begin{equation}
\label{g}
\sqrt{-g}g^{\mu\nu}= \sqrt{-\gamma}(\gamma^{\mu\nu} + h^{\mu\nu})~, 
\end{equation}
and $g_{\mu \nu}$ according to the definition 
$g_{\mu\alpha} g^{\alpha \nu}= \delta_{\mu}^{\nu}$. Then, the field 
equations (\ref{evac}), (\ref{emat}) absorb $\gamma^{\mu \nu}$
and $h^{\mu \nu}$ into $g^{\mu \nu}$ and take the form of Einstein's
geometrical equations 
\begin{equation}
R_{\mu\nu} = 0,~~~~~~~   R_{\mu\nu} - \frac1{2}g_{\mu\nu}R = \kappa T_{\mu\nu}, 
\label{1Gmn}
\end{equation}
where $R_{\mu\nu}$ is the Ricci curvature tensor constructed from 
$g_{\mu\nu}$ in the usual manner. The $g_{\mu\nu}$ can be interpreted 
as a metric tensor of some effective curved space-time. The matter 
energy-momentum tensor $T_{\mu\nu}$ in Eq.(\ref{1Gmn}) is variational 
derivative of $L^{m}$ with respect to $g_{\mu\nu}$, in contrast to 
$\tau^{\mu\nu}$ which was variational derivative of $L^{m}$ with respect 
to $\gamma_{\mu \nu}$.  

Note that the field equations as a whole can be rewritten in terms of 
$g_{\mu\nu}$, but not individual parts of these equations. The 
gravitational energy-momentum tensor $t^{\mu\nu}$ cannot be rewritten as a 
function of the tensor $g_{\mu\nu}$ and its first derivatives alone. This
is something to be expected, as there is no tensor that one could build 
from $g_{\mu\nu}$ 
and first derivatives of $g_{\mu\nu}$, apart of $g_{\mu\nu}$ itself. 
Therefore, there does not exist any meaningful gravitational 
energy-momentum tensor in the geometrical version of general relativity.

The universal character of the gravitational field, which allows one to `glue 
together' $\gamma^{\mu \nu}$ and $h^{\mu \nu}$ into a single object
$g^{\mu\nu}$ in the total Lagrangian and in the equations of motion, makes the 
flat space-time `non-observable' in the presence of gravitational fields. This 
is not surprising. We would have encountered the same `non-observability' of 
flat space-time in classical electrodynamics had we had access only to test 
particles with one and the same charge to mass ratio $e/m$. In the absence of 
neutral particles, which are capable of drawing the lines and angles of the 
Minkowski world in the region occupied by electromagnetic field, one would 
be given the option to interpret the motion of charged particles as arising 
due to the `curvature of space-time itself' rather than due to the 
external electromagnetic field. This is a possible interpretation, but not 
particularly illuminating, at least in this case.  

\begin{figure*}
\begin{center}
\includegraphics[height=6cm]{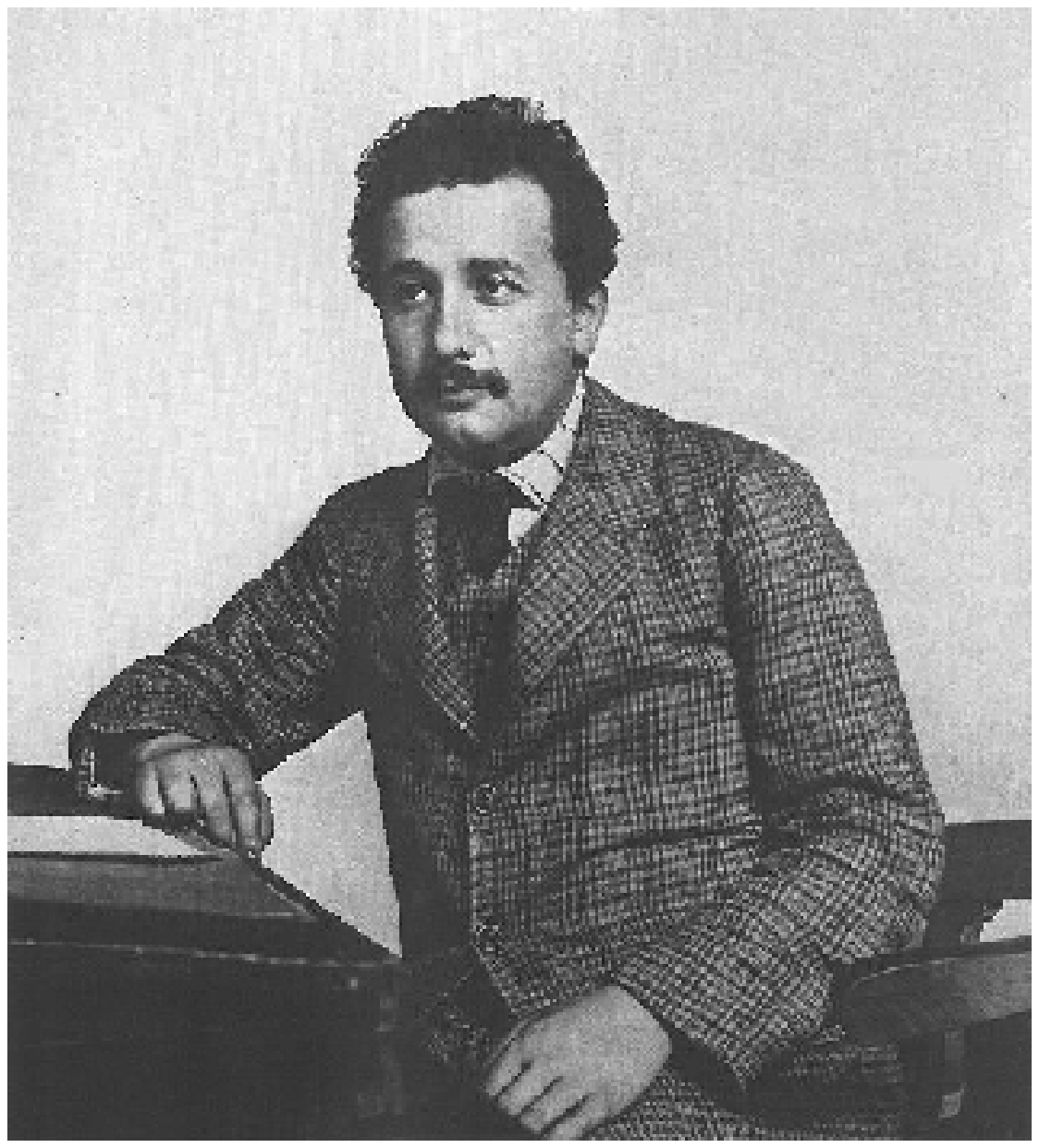} 
\includegraphics[height=6cm]{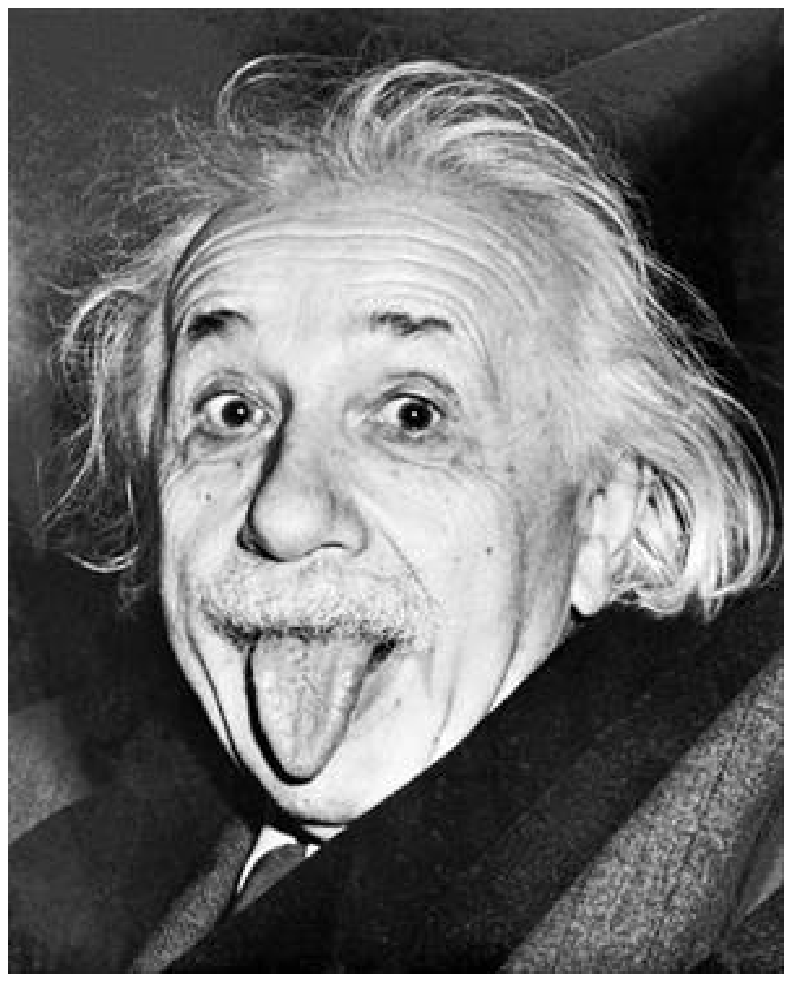}
\end{center}
\caption{Two famous photos of Einstein. Left: 1905, Bern.
Right: 1951, Princeton.}
\label{einst}
\end{figure*}

The generally covariant theories, including general relativity, admit arbitrary 
coordinate transformations. A coordinate transformation acting on an object, 
such for example as the metric tensor or the energy-momentum tensor of the 
electromagnetic field, can always be rearranged to state that the coordinates 
are not touched, but the object itself receives an increment in the same 
point. The rule of changing the numerical values of the object to the new ones 
involves the technique of Lie derivatives and depends on the transformation 
properties of the object and a given coordinate transformation. If a
dynamical equation is formulated as equality to zero of some tensor composed 
of the underlying objects, the increments of this tensor vanish when the 
dynamical equation is satisfied. So, a solution of this dynamical equation 
translates into a new solution. This procedure of changing the values of 
objects in the same coordinate frame and generating new solutions can be 
called a gauge transformation.

The ability to make arbitrary coordinate transformations in the 
field-theoretical general relativity can be rearranged to state even more.
Namely, that the coordinate system $x^{\alpha}$ and the metric tensor 
$\gamma_{\mu\nu}(x^{\alpha})$ remain untouched, but the set of gravitational 
field variables $h^{\mu \nu}(x^{\alpha})$, as well as matter variables, 
change to another set of variables in the same coordinate 
system $x^{\alpha}$. Under this operation, the gravitational 
field equations (\ref{evac}), (\ref{emat}), as well as the
matter field equations and the conservation laws (\ref{cons}), transform 
into a combination of themselves, so that a solution of field equations 
translates into another solution. A coordinate transformation 
interpreted this way can be called a `true' gauge transformation, 
because it looks very much similar to the gauge (gradient) 
transformation of classical electrodynamics, which changes electromagnetic
potentials, but not coordinates and metric. However, the origin of gauge 
transformations in gravity is different - it is all the same arbitrary 
coordinate transformations. In general, there is no any other symmetries.

While the equations as a whole are gauge-invariant (transform into a 
combination of themselves) under the action of `true' gauge transformations, 
individual parts of the equations are not. For example, in Eq.(\ref{evac}) 
the l.h.s. and r.h.s. receive individual non-zero increments, but such that 
they precisely cancel each other, leaving Eq.(\ref{evac}) gauge-invariant. 
To demand the gauge-invariance of $t^{\mu\nu}$ on its own, thus attempting to
make $t^{\mu\nu}$ ``physically significant" and not ``devoid of physical 
meaning" [\cite{blh}], would be equivalent to demanding that the field 
equations should be violated after a gauge transformation. 

The same holds true for conservation laws (\ref{cons}). Even if 
the energy-momentum tensor $\tau^{\mu\nu}$ represents only a couple of free 
particles, $\tau^{\mu\nu}$ changes under the action of `true' gauge 
transformations. So, the associated change of $t^{\mu\nu}|_{m}$ must 
take care of new positions and 
new dynamics of the particles. The $t^{\mu\nu}|_{m}$ cannot be
gauge-invariant on its own, but the gauge-related solutions are 
observationally equivalent, at least in the classical domain of the 
theory, and as long as one ignores initial and boundary conditions. 

As far as one can see at present, the geometrical and field-theoretical 
pictures of gravity are representations of one and the same theory of general 
relativity, not different theories. Each of the viewpoints has its advantages 
and disadvantages, and we have to become eloquent in both of them. Feynman 
once remarked [\cite{f}]: ``if the peculiar viewpoint taken is truly 
experimentally equivalent to the usual in the realm of the known there
is always a range of applications and problems in this realm for which the
special viewpoint gives one a special power and clarity of thought, which
is valuable in itself".

In Fig.\ref{einst} one can see two famous photographs of Einstein. He is shown 
at times of the beginning and the end of his scientific career. In the left 
photo, the physicist Einstein is here, in Bern, in 1905. He was denouncing 
the notions of absolute space and time, but was not yet under the influence 
of geometrical techniques. In the right photo,
Einstein is in Princeton, in 1951, when the idea that the space-time is
something like a ``fabric", which can curve, wrap, expand, oscillate, etc.
was accepted. By that time, Einstein has twice changed his opinion about the
reality of gravitational waves. This second picture, is it not addressed to
those of us who believe too much literally that ``gravity is geometry" ?

\section{Generalising the General Relativity}
\label{sec:6}

The special power and clarity of thought mentioned by Feynman may be 
especially valuable now, when we are facing the situation that some 
modifications of general relativity may be required. Hopefully, the recent 
indications on the accelerated expansion of our approximated homogeneous 
isotropic Universe can be resolved by more accurate understanding of the 
limits of applicability of this approximation, and this will be so much
for the ``dark energy".  But if not, the prospect of modifications of 
Einstein's gravity will be looming large. The internal consistency of the
candidate modified theory will be a larger hurdle to overcome than agreement
with observations.    

Without a physical guidance, we are in front of an ocean of possibilities. 
The geometrical route ``gravity is geometry" will definitely lead to 
undesirable higher-order differential equations in terms of $g_{\mu\nu}$, 
as there is no way of modifying the second-order dynamical equations 
(\ref{1Gmn}) except adding a new fundamental constant - the cosmological 
$\Lambda$-term. In contrast, the very logic of the field-theoretical 
approach allows natural and consistent generalisations of general relativity 
without raising the order of differential field equations [\cite{bg}]. These 
generalisations are based on the possibility (and maybe necessity) of adding 
the `mass'-term   
\begin{equation}
\label{two}
\sqrt{-\gamma}\left[ k_1h^{\rho\sigma}h_{\rho\sigma} + k_2 h^2\right]
\end{equation} 
to the gravitational Lagrangian $L^{g}$. In this expression, the 
constants $k_1$ and $k_2$ have dimensionality of $[length]^{-2}$, and 
$h \equiv h^{\mu\nu} \gamma_{\mu\nu}$. 

\begin{figure}
\includegraphics[width=9cm,height=7cm]{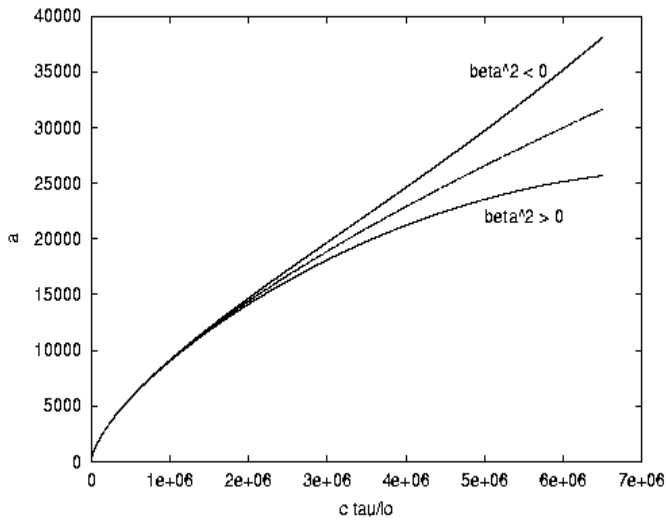}
\caption{The dashed line is a Friedmann solution of general relativity.
The upper solid line is a solution of the finite-range gravity for
$\left(\frac{cm_0}{\hbar}\right)^2 \equiv \beta^2 < 0$, while the lower 
solid line is a solution for $\beta^2 > 0$.}
\label{cosmol}
\end{figure}

The linearised approximation of this generalised gravity allows one to 
interpret the introduced constants as masses $m_2$ and $m_0$ of $spin-2$ 
and $spin-0$ gravitons,
\begin{equation}
\label{mm}
\left(\frac{cm_2}{\hbar}\right)^2 =  4k_1, ~~~~~~
\left(\frac{cm_0}{\hbar}\right)^2 =-2k_1 \frac{k_1+4k_2}{k_1+k_2}.
\end{equation}
The interpretation in terms of masses implies $m_0 >0$ and $m_2 >0$, but 
the theory allows also $(m_0)^2 <0$ and $(m_2)^2 <0$. When both constants
$(m_0)^2$ and $(m_2)^2$ are sent to zero, this finite-range gravitational 
theory smoothly approaches the equations and observational predictions of
the massless general relativity - the property not shared by many other
proposed modifications. 

It is amazing to see how much the class of allowed solutions broadens and 
changes under the impact of the seemingly innocent modification (\ref{two}) 
of general relativity. The modifications affect the polarization states 
and propagation of gravitational waves, the event horizons of black holes, 
the early-time and late-time cosmological evolution. In cosmological 
applications, the signs and values of $(m_0)^2$ and $(m_2)^2$ are crucial 
for the character of changes in the early-time and late-time 
evolution. In particular, if one allows $(m_0)^2$ to be negative, the 
late-time evolution of a homogeneous isotropic universe experiences an 
accelerated expansion, which we may be required to explain if the 
observations keep demanding this phenomenon. The graph of the modified 
scale factor $a$ as a function of time is shown in Fig.\ref{cosmol}, 
taken from the second paper in Ref.\cite{bg}.

\section{Conclusions}
\label{sec:7}

The quest for better understanding of gravity is far from over. This 
conference is certainly not the last one on the topic of nature of gravity.
Hopefully, important progress will be reached in the near future, 
particularly as a result of new experiments, such as $Planck$ and others. 
Having discovered relic gravitational waves, we will be more confident about 
the limits of applicability of general relativity and will better understand 
how cosmic gravity works in concert with quantum mechanics. More excitment 
is anticipated.   

\begin{acknowledgements}
I am grateful to D.Baskaran and W.Zhao for help in preparation of this 
manuscript.
\end{acknowledgements}

\end{document}